\documentclass{article}

\usepackage[english]{babel}
\usepackage[letterpaper,top=2cm,bottom=2cm,left=2.5cm,right=2.5cm,marginparwidth=1.75cm]{geometry}
\usepackage{footmisc}
\usepackage{appendix}
\usepackage{amsmath}
\usepackage{float}
\usepackage{graphbox}
\usepackage{authblk}
\usepackage[colorlinks=true, allcolors=blue]{hyperref}
\usepackage{subcaption}
\usepackage{indentfirst}
\usepackage{graphicx}
\usepackage{comment}
\usepackage{multirow}
\usepackage{xcolor}
\usepackage[figurename=Fig.]{caption}
\usepackage{cite}
\usepackage{amsfonts}

\usepackage{booktabs}
\usepackage{siunitx}

\title{Optimizing Quantum Convolutional Neural Network Architectures for Arbitrary Data Dimension}
\author[1]{\normalsize Changwon Lee}
\author[1]{\normalsize Israel F. Araujo}
\author[2,$\dagger$]{\normalsize Dongha Kim}
\author[3,$\dagger$]{\normalsize Junghan Lee}
\author[4,$\dagger$]{\normalsize Siheon Park}
\author[2,5,$\dagger$]{\normalsize Ju-Young Ryu}
\author[1,6,*]{\normalsize Daniel K. Park}
\affil[1]{\small \textit{Department of Statistics and Data Science, Yonsei University, Seoul, Republic of Korea}}
\affil[2]{\small \textit{School of Electrical Engineering, KAIST, Daejeon, Republic of Korea}}
\affil[3]{\small \textit{Department of Physics, Yonsei University, Seoul, Republic of Korea}}
\affil[4]{\small \textit{Department of Physics \& Astronomy, Seoul National University, Seoul, Republic of Korea}}
\affil[5]{\small \textit{Quantum Computing R\&D, Norma Inc., Seoul, Republic of Korea}}
\affil[6]{\small \textit{Department of Applied Statistics, Yonsei University, Seoul, Republic of Korea}}

\date{}

\begin{document}
\maketitle
\footnotetext{\small * : corresponding author}
\footnotetext{\small Email addresses : dkd.park@yonsei.ac.kr}
\footnotetext{\small $^{\dagger}$These authors contributed equally to this research and listed in alphabetical order.}
\vspace{-8mm}
\begin{abstract}
Quantum convolutional neural networks (QCNNs) represent a promising approach in quantum machine learning, paving new directions for both quantum and classical data analysis. This approach is particularly attractive due to the absence of the barren plateau problem, a fundamental challenge in training quantum neural networks (QNNs), and its feasibility. However, a limitation arises when applying QCNNs to classical data. The network architecture is most natural when the number of input qubits is a power of two, as this number is reduced by a factor of two in each pooling layer. The number of input qubits determines the dimensions (i.e. the number of features) of the input data that can be processed, restricting the applicability of QCNN algorithms to real-world data. To address this issue, we propose a QCNN architecture capable of handling arbitrary input data dimensions while optimizing the allocation of quantum resources such as ancillary qubits and quantum gates. This optimization is not only important for minimizing computational resources, but also essential in noisy intermediate-scale quantum (NISQ) computing, as the size of the quantum circuits that can be executed reliably is limited. Through numerical simulations, we benchmarked the classification performance of various QCNN architectures when handling arbitrary input data dimensions on the MNIST and Breast Cancer datasets. The results validate that the proposed QCNN architecture achieves excellent classification performance while utilizing a minimal resource overhead, providing an optimal solution when reliable quantum computation is constrained by noise and imperfections.
\end{abstract}

\section{Introduction}
\label{sec:introduction}
The advent of deep neural networks (DNNs) has transformed machine learning, drawing considerable research attention owing to the efficacy and broad applicability of DNNs ~\cite{jain1996artificial,vaswani2017attention}. Among the DNNs, convolutional neural networks (CNNs) have emerged to pivotally contribute toward image processing and vision tasks ~\cite{lecun1989backpropagation,lecun1998gradient}. By leveraging filtering techniques, the CNN architecture effectively detects and extracts spatial features from input data. CNNs exhibit exceptional performance in diverse domains---including image classification, object detection, face recognition, and medical image processing---and have attracted interest from both researchers and industry ~\cite{krizhevsky2012imagenet,he2016deep,girshick2014rich,li2015convolutional,tajbakhsh2016convolutional}.

Although DNNs have proven successful in various data analytics tasks, the increasing volume and complexity of datasets present a challenge to the current classical computing paradigm, prompting the exploration of alternative solutions. Quantum machine learning (QML) has emerged as a promising approach to address the fundamental limitations of classical machine learning. By leveraging the advantages of quantum computing techniques and algorithms, QML aims to overcome the inherent constraints of its classical counterparts~\cite{biamonte2017quantum,schuld2019quantum,schuld2015introduction,lloyd2013quantum}. However, a challenge in contemporary quantum computing lies in the difficulty of constructing quantum hardware. This challenge is characterized by noisy intermediate-scale quantum (NISQ) computing~\cite{preskill2018quantum,bharti2022noisy}, as the number of quantum processors that can be controlled reliably is limited owing to noise. Quantum-classical hybrid approaches based on parameterized quantum circuits (PQCs) have been developed to enhance the utility of NISQ devices~\cite{benedetti2019parameterized,sim2019expressibility,cerezo2021variational}. These strategies have contributed to advancements in quantum computing and machine learning, facilitating improved performance and applicability in various domains. In particular, PQC-based QML models have demonstrated a potential to outperform classical models in terms of sample complexity, generalization, and trainability ~\cite{huang2021information,aharonov2022quantum,schuld2021effect,caro2021encoding,thanasilp2208exponential,abbas2021power,caro2022generalization}. 
However, PQCs encounter a critical challenge in addressing real-world problems, particularly in relation to scalability, which is attributed to a phenomenon known as barren plateaus (BP)~\cite{mcclean2018barren}. This phenomenon is characterized by an intrinsic tradeoff between the expressibility and trainability of PQCs~\cite{PRXQuantum.3.010313}, causing the gradient of the cost function to vanish exponentially with the number of qubits under certain conditions. An effective strategy for avoiding BPs is to adopt a hierarchical quantum circuit structure, wherein the number of qubits decreases exponentially with the depth of the quantum circuit~\cite{grant2018hierarchical,pesah2021absence}. Quantum convolutional neural networks (QCNNs) notably employ this strategy, as highlighted in recent studies~\cite{cong2019quantum,hur2022quantum,kim2023classical,laurens2023qcnnnas,oh2023quantum}. Inspired by the CNN architecture, the QCNN is composed of a sequence of quantum convolutional and pooling layers. Each pooling layer typically reduces the number of qubits by a factor of two, thereby increasing the quantum circuit depth to $O(\log(n))$ for $n$ input qubits. The logarithmic circuit depth is one of the features that renders the QCNN an attractive architecture for NISQ devices, implying that the most natural design approach is to set the number of input qubits to a power of two. However, the number of input qubits required is determined by the input data dimension, i.e., the number of features in the data. If the input data require a number of qubits that is not a power of two, some layers will inevitably have odd numbers of qubits. This can occur either in the initial number of input qubits or during the pooling operation, representing a deviation from the optimal design and requiring appropriate adjustments. In particular, having an odd number of qubits in a quantum convolutional layer results in an increase in the circuit depth if all nearest-neighbor qubits interact with each other. Consequently, the run time increases and noise can negatively impact the overall performance and reliability of the QCNN.
Moreover, it is unclear how breaking translational invariance in the pooling layer, a key property of the QCNN, affects overall performance. Because these considerations constrain the applicability of the QCNN algorithm, our goal is to optimize the QCNN architecture, developing an effective QML algorithm capable of handling arbitrary data dimensions.
 
In this study, we propose an efficient QCNN architecture capable of handling arbitrary data dimensions. Two naive approaches served as baselines to benchmark the proposed architectures: the classical data padding method, which increases the input data dimension through zero padding or periodic padding to encode it as a power of two, and the skip pooling method, which directly passes one qubit from each layer containing an odd number of qubits to the next layer without pooling. The first method requires additional ancillary qubits without increasing the circuit depth, whereas the second method does not require ancillary qubits but results in an increased circuit depth to preserve the translational invariance in the convolutional layers. By contrast, our proposed method effectively optimizes the QCNN architecture by applying a qubit padding technique that leverages ancillary qubits. By introducing an ancillary qubit into layers with an odd number of qubits, we can effectively construct convolutional layers without an additional increase in circuit depth. This enables a reduction in the total number of qubits by up to $\log(n)$ compared with the classical data padding method. Moreover, the reuse of a single ancillary qubit across multiple layers further reduces the required number of ancillary qubits. This strategy of qubit reuse efficiently optimizes the number of ancillary qubits along with the circuit depth. In addition, recycling the ancilla qubit facilitates uniform operations across layers, systematically enhancing the stability and efficiency of the QCNN architecture. To validate our approach, we benchmarked our proposed method against naive methods on two datasets: MNIST and Breast Cancer. Numerical simulation results show that our proposed method achieves a high classification accuracy comparable to that of naive methods. Notably, our method significantly reduces the number of qubits used compared with classical data padding methods, providing substantial advantages in terms of resource efficiency. We also conducted noise simulations using information from an IBM quantum device that mimics the operations and characteristics of real quantum hardware. The noise simulation results demonstrate that the proposed method exhibits less performance degradation and lower variability under realistic noise conditions than the skip pooling method. This is a consequence of the skip pooling method requiring a larger circuit depth. Because the proposed method not only improves the runtime but also enhances robustness against noise, it serves as a fundamental building block for the effective applicability of QCNNs to real-world data with an arbitrary number of features.

The remainder of this paper is organized as follows. We introduce the foundational concepts of QML in Section~\ref{sec:section2}, focusing on principles underlying quantum neural networks (QNNs) and QCNNs. Section~\ref{sec:section3} presents the detailed design of a QCNN architecture capable of handling arbitrary data dimensions, including a comparative analysis between naive methods and our proposed methods. Simulation results are presented in Section~\ref{sec:section4} along with a comparative performance analysis of the naive and proposed methods in the classification of MNIST and Breast Cancer data under both noiseless and noisy conditions. Section~\ref{sec:section5} explores possible extensions of multi-qubit quantum convolutional operations. Finally, concluding remarks are presented in Section~\ref{sec:section6}.

\section{Background}
\label{sec:section2}
\begin{figure}[ht]
  \centering
  \includegraphics[width=\linewidth]{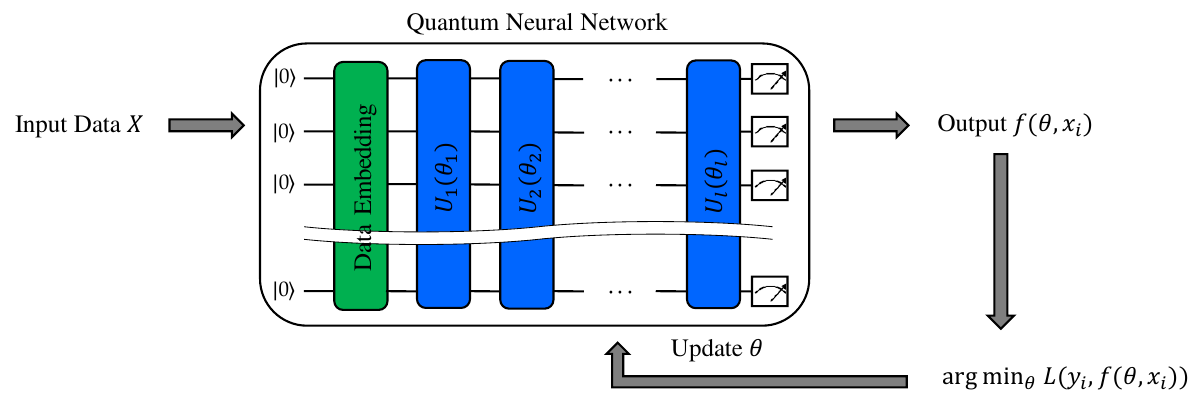}
  \caption{A schematic of the training process of a QNN. This figure outlines the sequential steps involved in training a QNN, starting with data preparation and qubit initialization, followed by the application of quantum gates to embed the data, and using a PQC for training. The output is obtained from measurements of the quantum state. Parameters are updated by minimizing the loss function.
  }
  \label{fig:fig1}
\end{figure}

\subsection{Quantum Neural Network}
A DNN is a machine learning model constructed by deeply stacking layers of neurons~\cite{Goodfellow-et-al-2016}. Using nonlinear activation functions---such as the sigmoid, ReLU, and hyperbolic tangent functions---the DNN can learn patterns in complex data to solve various problems with high performance. Although the mathematical foundation for the success of DNNs remains an active area of research~\cite{10.1145/3446776}, several studies, as well as the universal approximation theorem, have demonstrated that neural networks can approximate complex functions with arbitrary accuracy~\cite{lu2020universal}. On the other hand, a QNN is a quantum machine learning model, where the data is propagated through a PQC in the form of a quantum state. The data can be either intrinsically quantum, if the data source is a quantum system, or classical. In the latter case, which is the primary focus of this work, the classical data first has to be mapped to a quantum state. Note that nonlinear transformation of the input data can occur during this data mapping step. Since the parameters of the PQC are real-valued and its output is differentiable with respect to the parameters, they are typically trained through classical optimizers, similar to how DNNs are trained. In this sense, the QNN-based ML is also known to be a quantum-classical hybrid approach.
Quantum-classical hybrid approaches using PQCs are effective at shallow circuit depths~\cite{cerezo2021variational}, which significantly enhances their applicability to NISQ devices with limited numbers of qubits. In addition, the PQC can approximate a broad family of functions with arbitrary accuracy, making it a good machine learning model~\cite{benedetti2019parameterized,schuld2021effect}.  

A QNN consists of three primary components: (1) Data Embedding, (2) Data Processing, and (3) Measurements. These models transform classical data into quantum states, to be processed using a sequence of parameterized quantum gates. The training process connects the measurement results to the loss function, which is used to tune and train the parameters. Figure~\ref{fig:fig1} depicts the overall training process of a QNN. Consider a dataset $\mathcal{D} = \left\{ (\boldsymbol{x}_{i}, y_{i}) \right\}_{i = 1}^{M}$ with $\boldsymbol{x}_i\in\mathbb{R}^{N}$ and $y_i\in\mathbb{R}$, and PQC $U_{i}(\theta_{i})$, where $\theta_{i} = (\theta^{1}, \ldots, \theta^{m})$ represents a set of tunable parameters. Typically, the dataset is embedded into a quantum Hilbert space by a unitary transformation applied to $n$ qubits initially prepared in $| 0 \rangle^{\otimes n}$. Denoting the data embedded state as $| \psi_{\mathrm{in}} \rangle$, the final state of the QNN can be expressed as follows:
\begin{equation}
    | \psi_{\mathrm{out}} \rangle = U_{l}(\theta_{l}) U_{l-1}(\theta_{l-1}) \ldots U_{1}(\theta_{1}) | \psi_{\mathrm{in}} \rangle.
\end{equation} 
The output function of the QNN is $f(\boldsymbol{\theta}, \boldsymbol{x}_{i}) = \langle \psi_{\mathrm{out}} | \mathcal{O} | \psi_{\mathrm{out}} \rangle $, where $\mathcal{O}$ is an observable of the quantum circuit. The parameters are optimized using classical methods such as the gradient descent algorithm ~\cite{ruder2016overview}, which minimizes the following loss function:
\begin{equation}
L(\boldsymbol{\theta}) = \frac{1}{M} \sum_{i=1}^M {| y_{i} - f(\boldsymbol{\theta}, \boldsymbol{x}_{i}) |^{2}}.
\end{equation}
Furthermore, gradients in quantum computing can be computed directly using methods such as parameter-shift rules, wherein derivatives are approximated by shifting the parameters at fixed intervals and then measuring the difference in the output of the quantum circuit as a result of that change ~\cite{mitarai2018quantum,schuld2019evaluating}. However, as the number of qubits in the training PQC increases, the parameter space of the quantum circuit also increases, leading to the BP phenomenon ~\cite{mcclean2018barren}. Although this phenomenon represents a significant performance limitation of QML, it can be mitigated by applying hierarchical structures in the quantum circuits~\cite{grant2018hierarchical}, such as the QCNN~\cite{pesah2021absence}.

\subsection{Quantum Convolutional Neural Network}
The QCNN is a type of PQC inspired by the concept of CNNs. QCNNs exhibit the property of translational invariance, with quantum circuits sharing the same parameters within the convolutional layer, and reduce dimensionality by tracing out some qubits during the pooling operation. A primary distinction between a QCNN and a CNN is that data in a QCNN are defined in a Hilbert space that grows exponentially with the number of qubits. Consequently, whereas classical convolution operations typically transform vectors into scalars, quantum convolution operations perform more complex linear mapping, transforming vectors into vectors through a unitary transformation of the state vector. Thus, quantum convolutional operations are distinct from classical convolutional operations. Problems defined in the exponentially large Hilbert space are intractable in a classical setting; however, QCNNs offer the possibility of effectively overcoming these challenges by utilizing qubits in a quantum setting. QCNNs have also demonstrated the capability to classify images in a manner similar to their classical counterparts~\cite{hur2022quantum}. In the convolutional layer, local features are extracted through unitary single-qubit rotations and entanglements between adjacent qubits, and the features dimensions are reduced in a pooling layer. Typically, the pooling layer includes parameterized two-qubit controlled unitary gates, and the control qubit is traced out after the gate operation to halve it. In binary classification tasks (i.e. $y_i\in\lbrace +1,-1\rbrace$), a QCNN repeats the convolutional and pooling layers until only one qubit remains, and performs classification by measuring the last qubit. These architectures maintain a shallow circuit depth by effectively reducing the number of qubits through a hierarchical structure, which is crucial for improving model performance and avoiding BP. In addition, by turning off the translational invariance property that sharing the same parameters within the convolutional or pooling layer, more parameters can be introduced to the QML model while preserving the absence of BP. The structure between layers ensures a circuit depth of $O(\log(n))$ for $n$ input qubits. In particular, the shallow depth of the QCNN contributes to its high performance in NISQ devices. In addition, the simple structure of repetitive circuits allows for the applicability of QCNNs to a wide range of tasks including classical and quantum data classification~\cite{cong2019quantum,maccormack2022branching,hur2022quantum}, error correction~\cite{cong2019quantum}, and classical to quantum transfer learning~\cite{mari2020transfer,kim2023classical}.

\section{QCNN architectures for arbitrary data dimensions}
\label{sec:section3}
Data encoding is a crucial process in quantum computing that transforms classical data into quantum-state. During this process, the input data dimensions determine the number of qubits required to represent the quantum state. For example, amplitude encoding allows the input data $\boldsymbol{x} = (x_{1}, \ldots, x_{N})^{T}\in\mathbb{R}^{N}$ with dimensions $N = 2^{n}$ to be represented as amplitudes of an $n$-qubit quantum state. Since a pooling layer in QCNN discards half of the qubits, $n$ also has to be a power of two. The condition requiring the number of input qubits to be a power of two plays a critical role in the efficient applicability of QCNN algorithms. However, the dimensions of classical data do not always conform to this condition. In this section, we describe our proposed method that enables QCNN architectures to handle arbitrary data dimensions, along with naive baseline methods. To demonstrate the efficiency of the proposed method, we analyzed the number of ancillary qubits, parameters, and circuit depths when implementing the QCNN algorithm. 

\subsection{Naive Methods}
\subsubsection{Classical data padding}
\begin{figure}[ht]
  \centering
  \includegraphics[width=\linewidth]{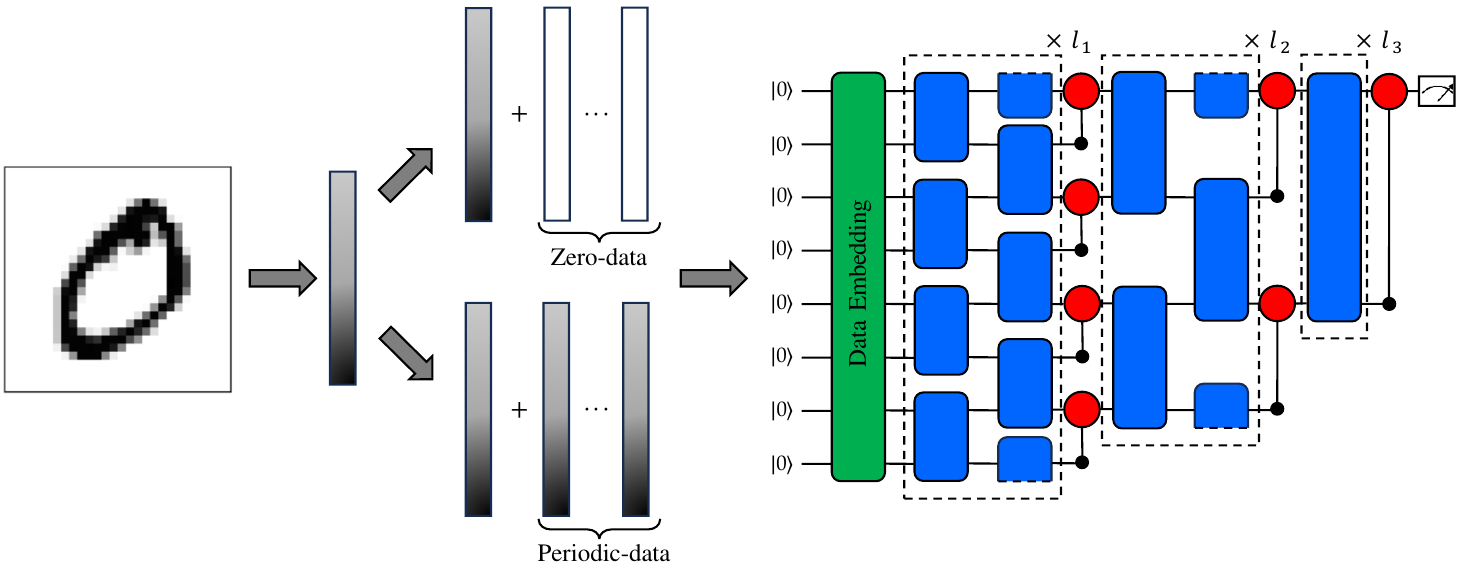}
  \caption{
  Schematic of the QCNN algorithm with eight input qubits using the classical data padding method. The quantum circuit consists of three components: data embedding (green squares), convolutional gate (blue squares), and pooling gate (red circles). The data embedding component is further divided into two methods: top embedding, which pads the zero-data, and bottom embedding, which pads the input data repeatedly. The convolutional and pooling gate use a PQC. Throughout the hierarchy, the convolutional gate consistently applies the same two-qubit ansatz to the nearest-neighboring qubit in each layer. In the $i$th convolutional layer, the set of gates that completes a loop connecting all nearest-neighboring qubits and the qubits at the boundaries can be repeated $l_{i}$ times. The pooling gate uses the same approach, and can be represented as a controlled unitary transformation that is activated when the control qubit is 1.
  }
  \label{fig:fig2}
\end{figure}
In a CNN, the direct application of kernels to input feature maps during convolution operations can reduce the output feature map size compared to that of the input, leading to the potential loss of important information. Padding techniques that artificially enlarge the input feature map by adding specific values (typically zeros) around the input data are employed to prevent information loss and enhance model training. Inspired by CNN, similar padding strategies can be applied in QCNNs by increasing the data dimensions until the data can be encoded into a number of input qubits with a power of two, by either adding a constant value (zero) or periodically repeating the input data. These methods are referred to as `zero-data padding' and `periodic-data padding', respectively. Figure~\ref{fig:fig2} illustrates an example of the classical data padding method, with a handwritten digit image reduced to 30 dimensions. This 30-dimensional input can be encoded into five input qubits using amplitude encoding. Classical data padding methods can be applied to expand the data dimension, aligning the number of input qubits to a power of two. By using either zero- or periodic-data padding, the data dimensions can be expanded to $2^8$. Through these padding procedures, three additional ancillary qubits are introduced to the original system of five input qubits, resulting in an eight-qubit QCNN structure. Classical data padding approaches not only increase the number of qubits but may also result in poor classification performance because the number of dummy features that are added can often be significantly larger than the number of data features.

We denote the initial number of input qubits as $K$, and let $\lceil \log_2(K)\rceil = m$. Classical data padding typically employs $2^{m} - K$ ancillary qubits. Without loss of generality, we assume that both the convolutional and pooling gates of the QCNN have one parameter and a depth of 1. The quantum circuit depth is $\sum_{i=1}^{m} (2 l_{i} + 1) - l_{m}$, where $l_{i}$ denotes the number of complete sets of two-qubit gates connecting the nearest-neighboring qubits as well as the top and bottom qubits in the $i$th convolutional layer as depicted in Fig.~\ref{fig:fig2}. If the parameters are shared, then the total number of parameters is $\sum_{i=1}^{m} (l_{i} + 1)$. However, if the parameters are not shared, then the total number of parameters is $\sum_{i=1}^{m-1} (\lceil {{2^{m}} \over {2^{i-1}}} \rceil (l_{i} + {1 \over 2} ) )+ l_{m} + 1$.

\subsubsection{Skip pooling}
\begin{figure}[ht]
  \centering
    \begin{subfigure}[b]{0.33\textwidth}
        \includegraphics[width=1\linewidth]{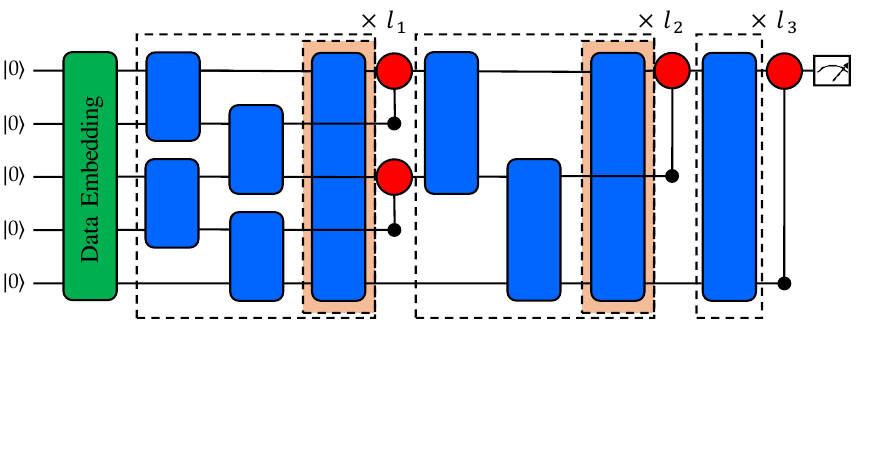}
        \caption{Skip pooling}
        \label{fig:fig3_1}
    \end{subfigure}%
    \begin{subfigure}[b]{0.33\textwidth}
        \includegraphics[width=1\linewidth]{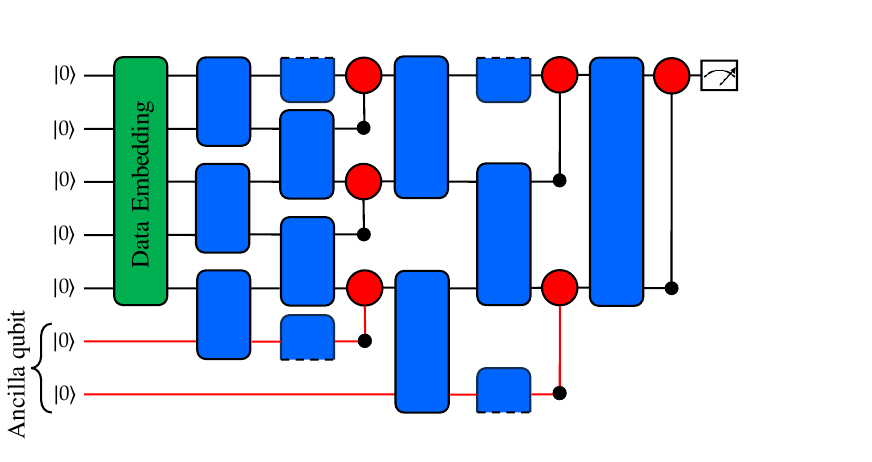}
        \caption{Layer-wise qubit padding}
        \label{fig:fig3_2}
    \end{subfigure}%
    \begin{subfigure}[b]{0.33\textwidth}
        \includegraphics[width=1\linewidth]{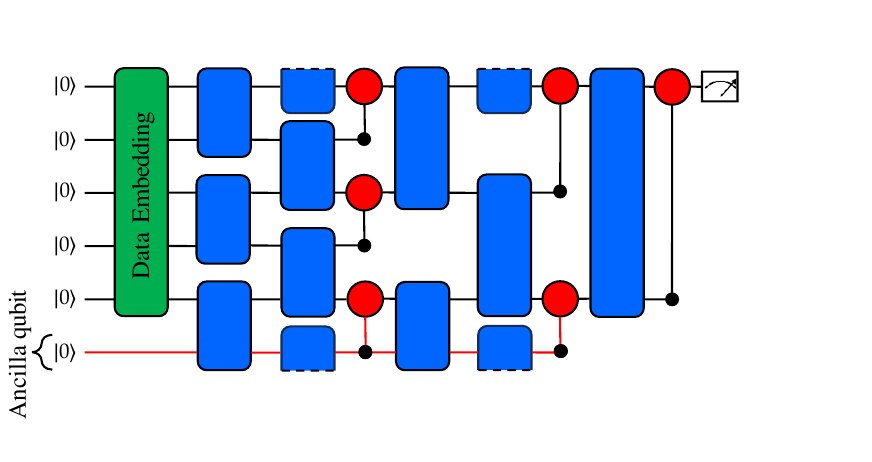}
        \caption{Single-ancilla qubit padding}
        \label{fig:fig3_3}
    \end{subfigure}%
  \caption{
  Schematic of a QCNN algorithm with five initial input qubits in a circuit with and without ancillary qubits. (a) uses a method called skip pooling to perform convolution and pooling operations between each qubit without ancillary qubits. (b) uses two ancillary qubits to construct the QCNN in a method called layer-wise qubit padding. The first layer has five qubits. Because this is an odd number of layers, one ancillary qubit is used to perform convolution and pooling operations. The second layer has three qubits, and another ancillary qubit is used. (c) uses only one ancillary qubit to construct the QCNN using a method called single-ancilla qubit padding. Unlike (b), the single ancillary qubit performs the convolution and pooling operations sequentially.
  }
  \label{fig:pad2}
\end{figure}

An alternate method enables implementation of the QCNN algorithm without the use of additional ancillary qubits. This method adopts a strategy where, within the QCNN structure, a qubit from each layer containing an odd number of qubits is passed directly to the next layer without performing a pooling operation. Although this approach minimizes the use of qubits, it inherently increases the circuit depth during convolutional operations in layers with odd numbers of qubits. This can affect the overall efficiency and execution speed of the quantum circuits. We refer to this method as `skip pooling'. Figure~\ref{fig:fig3_1} depicts an example of skip pooling, with a 30-dimensional input encoded into five input qubits using amplitude encoding. In the first layer of the QCNN, the convolutional operation between neighboring qubits introduces one more gate over classical data padding. 
Then, the $5\mathrm{th}$ qubit passes directly to the next layer without any pooling operations. This procedure is repeated for the second layer. When applied to layers that contain an odd number of qubits, skip pooling increases circuit depth during the convolutional operation. The increased circuit depth can be affected by noise, which may potentially propagate through each layer, reducing the accuracy and reliability of the information. This directly affects the efficiency and performance of the QCNN, necessitating an effective optimization strategy. 

Unlike classical data padding, no ancillary qubits are used; however, an additional circuit depth of $\sum_{i=1}^{m-1} Y_{i} l_{i}$ is incurred, where $Y_{i} := \lceil {{K} \over {2^{i-1}}} \rceil \mod 2$ is an odd number of qubits in the $i\mathrm{th}$ layer divided by the power of two. If the parameters are shared, then the total number of parameters is $\sum_{i=1}^{m} (l_{i} + 1)$. In contrast, if the parameters are not shared, then the total number of parameters is $\sum_{i=1}^{m-1} (\lceil {{K} \over {2^{i-1}}} \rceil (l_{i} + {1 \over 2} ) - {1 \over 2} Y_{i} ) + l_{m} + 1$. 

\subsection{Proposed Methods}

\subsubsection{Layer-wise qubit padding}
As an alternative to the aforementioned naive methods, we introduce a qubit padding method that leverages ancillary qubits in the QCNN algorithm. Whereas classical data padding requires additional ancillary qubits by increasing the size of the input data, qubit padding directly leverages ancillary qubits in the convolutional and pooling operations of the QCNN. Using ancillary qubits for layers containing odd numbers of qubits, we optimized the QCNN algorithm and designed an architecture capable of handling arbitrary data dimensions. We refer to this method as `layer-wise qubit padding'. Figure~\ref{fig:fig3_2} depicts an example of layer-wise qubit padding. As in the skip pooling example, a 30-dimensional input was encoded into five input qubits using amplitude encoding. In the first layer of the QCNN, one ancillary qubit is added to perform convolutional and pooling operations with neighboring qubits. This ensures pairwise matching between all qubits in two steps, avoiding additional circuit depth that may arise in skip pooling. This procedure is repeated for the second layer. Consequently, in a five-qubit QCNN with layer-wise qubit padding and two layers containing an odd number of qubits, two ancillary qubits are used to optimize the architecture. 

Layer-wise qubit padding generally requires $\sum_{i=1}^{m-1} Y_{i}$ ancillary qubits. The quantum circuit depth of $\sum_{i=1}^{m} (2 l_{i} + 1) - l_{m}$ is identical to that used in classical data padding. If the parameters are shared, then the total number of parameters is $\sum_{i=1}^{m} (l_{i} + 1)$. On the other hand, if the parameters are not shared, then the total number of parameters is $\sum_{i=1}^{m-1} ( (\lceil {{K} \over {2^{i-1}}} \rceil + Y_{i} ) (l_{i} + {1 \over 2} )) + l_{m} + 1 $, which is $\sum_{i=1}^{m-1} (Y_{i}) (l_{i} + 1)$ more than that for skip pooling. 

\subsubsection{Single-ancilla qubit padding}
Finally, we propose `single-ancilla qubit padding,’ a QCNN architecture designed to handle arbitrary input data dimensions using only one ancillary qubit. By reusing the ancillary qubit throughout the QCNN architecture, we significantly reduced the number of total qubits required for optimization. Figure~\ref{fig:fig3_3} illustrates an example of single-ancilla qubit padding. Unlike layer-wise qubit padding, this method reuses the same ancillary qubit for every layer with an odd number of qubits. Preserving the information of the ancillary qubit without resetting it when it is passed to the next layer plays a crucial role in enhancing the stability and performance of model training. 

\begin{figure}[ht]
  \centering
    \begin{subfigure}[b]{0.5\textwidth}
        \includegraphics[width=1\linewidth]{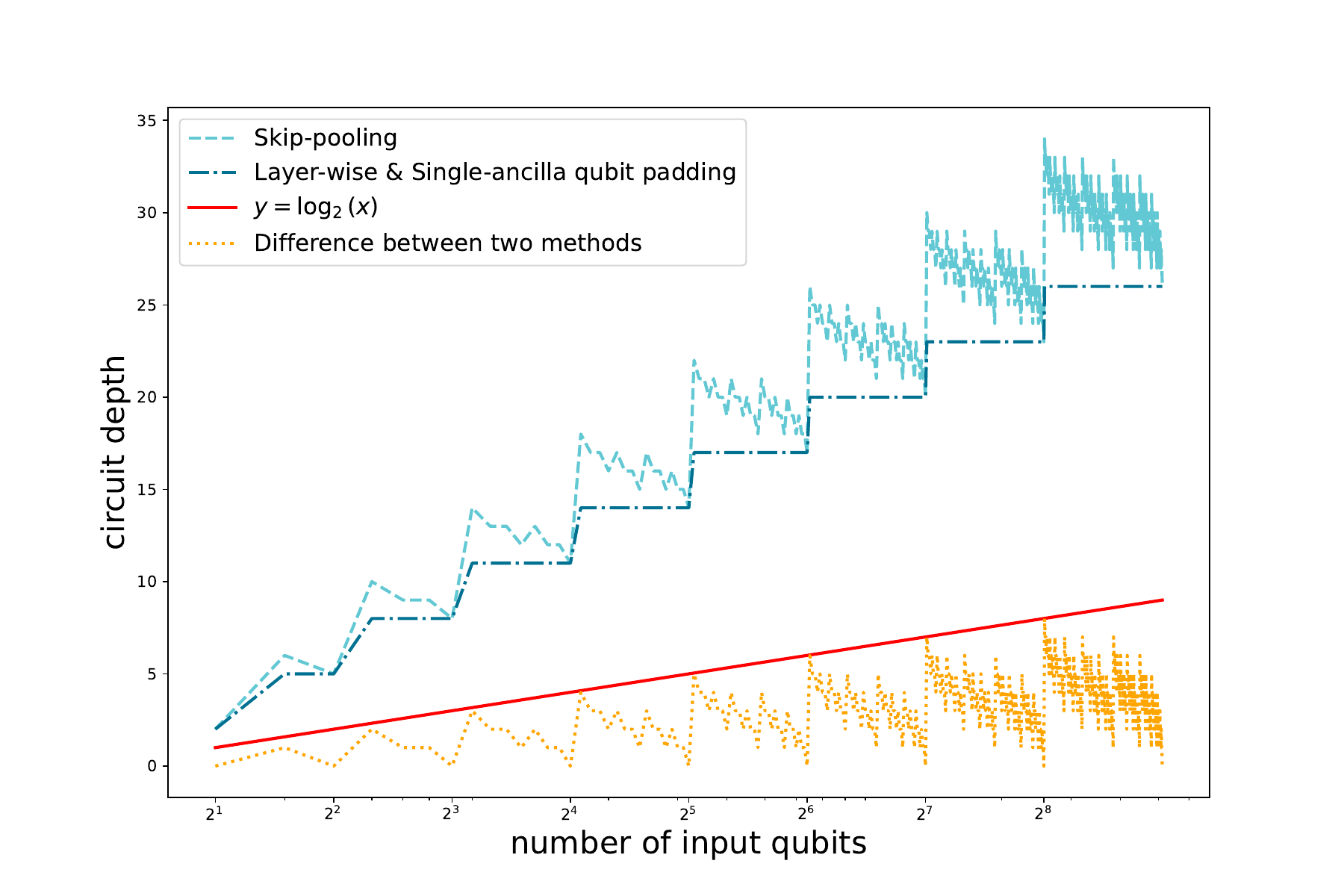}
        \caption{}
        \label{fig:fig4_1}
    \end{subfigure}%
    \begin{subfigure}[b]{0.5\textwidth}
        \includegraphics[width=1\linewidth]{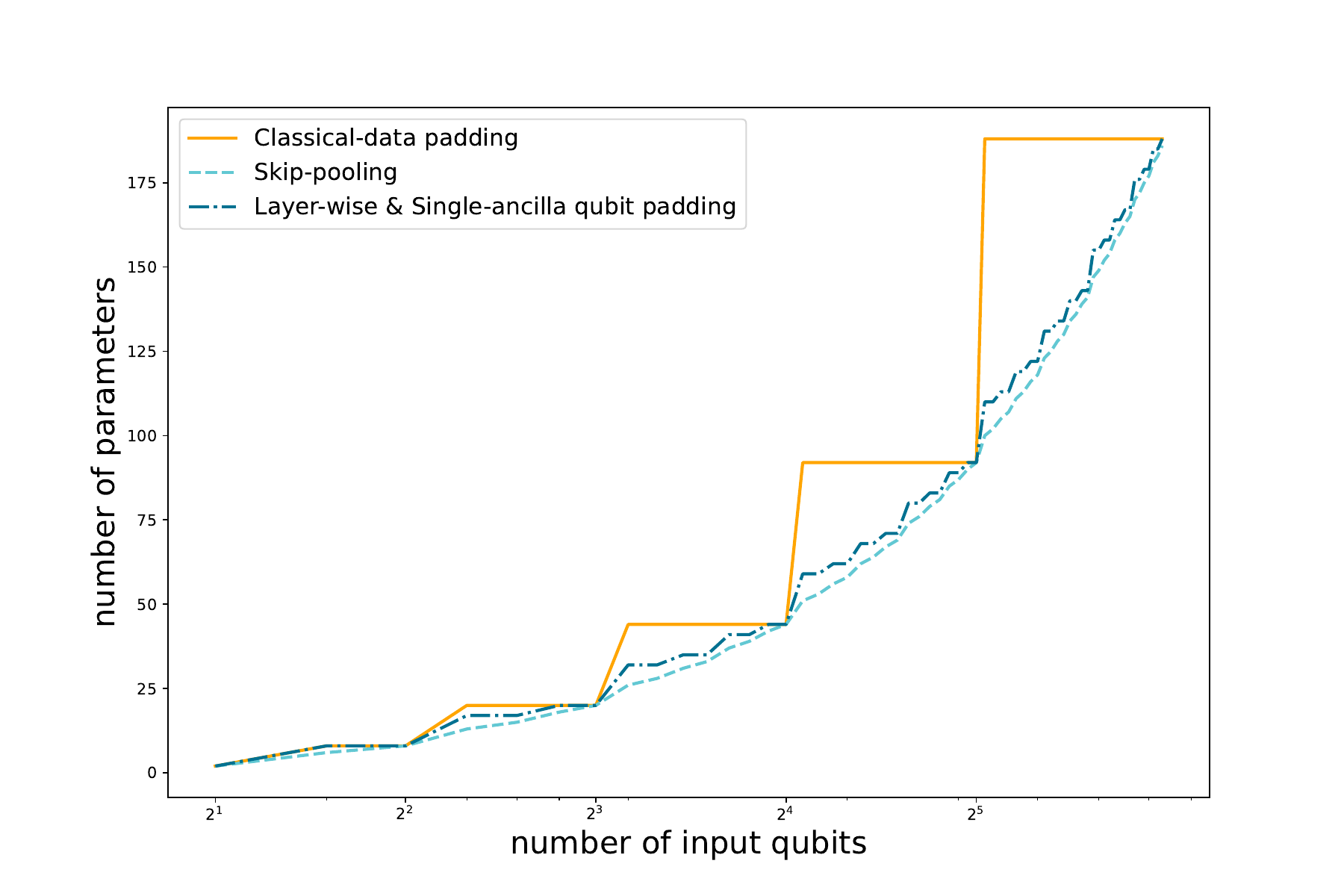}
        \caption{}
        \label{fig:fig4_2}
    \end{subfigure}%
    \caption{(a) Semi-log plot illustrating the difference in circuit depth between the skip pooling method and layer-wise \& single-ancilla qubit padding method. The dashed line represents circuit depth in the skip pooling method, the dash-dot line denotes circuit depth in the layer-wise \& single-ancilla qubit padding method, the dotted line represents the difference between the two methods, and the solid line corresponds to $\log_{2} x$, provided as a guide to the eye. (b) Semi-log plot illustrating the number of parameters in the case of parameter sharing off for the classical data padding, skip pooling, and layer-wise \& single-ancilla qubit padding methods. The solid line denotes the number of parameters for classical data padding, the dashed line represents the number of parameters for the skip pooling method, and the dash-dot line corresponds to the number of parameters for the layer-wise \& single-ancilla qubit padding.
  }
  \label{fig:fig4}
\end{figure}

Although single-ancilla qubit padding uses only one ancillary qubit, the quantum circuit depth and number of parameters remain the same as those in layer-wise qubit padding. Figure~\ref{fig:fig4_1} illustrates the circuit depths of the skip pooling and qubit padding methods. As the number of input qubits increases, the circuit depth of the qubit padding method is logarithmically less than that of the skip pooling method. Figure~\ref{fig:fig4_2} illustrates the number of parameters in the case of parameter-sharing off for the classical data padding, skip pooling, and layer-wise \& single-ancilla qubit padding methods. In the case of parameter-sharing on, the number of parameters is the same across all methods, hence we do not track how the number of parameters changes. We only compared the number of parameters in the case of parameter-sharing off. Without loss of generality, we assumed the convolutional layer $l_{i}$ to be equal to 1. When the number of input qubits is not a power of two, classical data padding uses the largest number of the parameters, whereas skip pooling and qubit padding use relatively fewer parameters. By introducing additional qubits into certain layers, the qubit padding method uses slightly more parameters than skip pooling. Therefore, single-ancilla qubit padding enables the design of efficient QCNN architectures with optimal allocation of quantum resources such as ancillary qubits and quantum gates. 

\section{Results}
\label{sec:section4}
The previous section provided an overview of various QCNN padding methods. In this section, we benchmark and evaluate our proposed padding methods in comparison with naive methods using a variety of classical datasets. To address the characteristics of NISQ devices, we added noise to the quantum circuits for benchmarking. 

\subsection{Methods and setup}
\subsubsection{Datasets}
Our experiments were conducted using the MNIST and Breast Cancer datasets. The MNIST dataset consists of handwritten digits, each represented as a $28 \times 28$ pixel image in grayscale. The dataset consists of a total of 60,000 training and 10,000 test images, each labeled with a numerical value ranging from 0 to 9. In our benchmarking, we focused on binary classification tasks by selecting two distinct pairs of labels: 0 \& 1 and 5 \& 6. In the noiseless scenario, we split the dataset into 10,000 training, 1,000 validation, and 1,000 test sets. In the noisy scenario, the dataset was halved. Furthermore, $28 \times 28$ features were relatively high-dimensional for current quantum hardware; therefore, we used principal component analysis as a dimensionality reduction technique to reduce the dataset to 30 features.

The Breast Cancer dataset classifies tumors as malignant or benign based on various attributes such as size and shape. The dataset contains a total of 569 instances with 30 features labeled as malignant or benign. We divided the data between 400 training, 75 validation, and 94 test sets. Furthermore, we applied kernel density estimation (KDE) to preprocess the dataset. KDE, a statistical method that estimates the probability density function of data using kernel functions, is useful when smoothly approximating features and interpreting the overall probabilistic characteristics. We preprocessed the features with a Gaussian kernel to enhance the effectiveness of model learning. 

\subsubsection{Ansatz}

\begin{figure}[ht]
  \centering
    \begin{subfigure}[b]{0.3\textwidth}
        \includegraphics{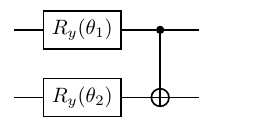}
        \caption{Convolutional circuit 1}
        \label{fig:fig5_1}
    \end{subfigure}%
    \begin{subfigure}[b]{0.3\textwidth}
        \includegraphics{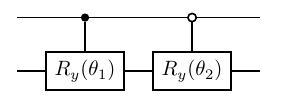}
        \caption{Pooling circuit}
        \label{fig:fig5_2}
    \end{subfigure}\\ %
    \begin{subfigure}[b]{0.9\textwidth}
        \includegraphics{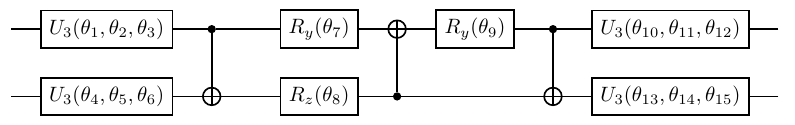}
        \caption{Convolutional circuit 2}
        \label{fig:fig5_3}
    \end{subfigure}%
    \caption{PQC for convolution and pooling operations. (a) and (b) are the convolutional and pooling circuits, respectively, that compose ansatz set1. (c) is the convolutional circuit that composes ansatz set2. $R_{i}(\theta)$ is the rotation by $\theta$ around the $i$-axis of the Bloch sphere, and $U_3(\theta, \phi, \lambda)$ is an arbitrary single-qubit gate, which can be expressed as $U_3(\theta, \phi, \lambda) = R_{z}(\phi) R_{x}(-\pi/2) R_{z}(\theta)R_{x}(\pi/2) R_{z}(\lambda)$.
    }
\end{figure}

We tested two different structures of the parameterized quantum circuit, also referred to as the ansatz, for the convolutional operations. The first one consists of two parameterized single-qubit rotations and a CNOT gate, as shown in Figure~\ref{fig:fig5_1} ~\cite{grant2018hierarchical}. This represents the simplest two-qubit ansatz. The second one is designed to express an arbitrary two-qubit unitary transformation. In general, any two-qubit unitary gate in the SU(4) group can be decomposed using at most three CNOT gates and 15 elementary single-qubit gates~\cite{maccormack2022branching,vatan2004optimal}. The quantum circuits shown in Figure~\ref{fig:fig5_3} represent the parameterization of an arbitrary SU(4) gate. Figure~\ref{fig:fig5_2} shows the pooling circuit, where two controlled rotations, $R_{y}(\theta_{1})$ and $R_{y}(\theta_{2})$, are applied, with each activated when the control qubit is 1 (filled circle) or 0 (open circle). Appendix~\ref{app:appendix_a} presents definitions of the quantum gates used in this study. We constructed two ansatz sets using a different combination of the convolutional and pooling circuits. Ansatz set 1 uses convolutional circuit 1 and a pooling circuit as shown in the figure, whereas ansatz set 2 uses convolution circuit 2 and pooling without a parameterized circuit. In the latter case, the pooling performs the partial trace operations without parameterized gates because the convolution circuit 1 is expressive enough to implement any two-qubit unitary operation. 

\subsection{Simulation without noise}

\begin{table}[h]
\centering
  \caption{Comparison of ancillary qubits, circuit depth, and parameters for classical data padding and skip pooling. $Y_{i} := \lceil {{K} \over {2^{i-1}}} \rceil \mod 2$ is an odd number of qubits in the $i\mathrm{th}$ layer when divided by a power of two.}
  \label{tab:table1}
  \begin{tabular}{ccc}
    \toprule
    Padding Methods & Classical data padding & Skip pooling \\
    \midrule
    Ancillary qubits & $2^{m} - K$ & 0 \\ 
    \midrule
    Circuit depth & $\sum_{i=1}^{m} (2 l_{i} + 1) - l_{m}$ & $\sum_{i=1}^{m} (2 l_{i} + 1) - l_{m}$ + $\sum_{i=1}^{m-1} Y_{i} l_{i}$ \\ 
    \midrule
    Parameter-sharing on & $\sum_{i=1}^{m} (l_{i} + 1)$ & $\sum_{i=1}^{m} (l_{i} + 1)$ \\ 
    \midrule
    Parameter-sharing off & $\sum_{i=1}^{m-1} (\lceil {{2^{m}} \over {2^{i-1}}} \rceil (l_{i} + {1 \over 2} ) )+ l_{m} + 1$ & $\sum_{i=1}^{m-1} (\lceil {{K} \over {2^{i-1}}} \rceil (l_{i} + {1 \over 2} ) - {1 \over 2} Y_{i} ) + l_{m} + 1$ \\ 
  \bottomrule
\end{tabular}
\end{table}

\begin{table}[h]
\centering
  \caption{Comparison of ancillary qubits, circuit depth, and parameters for layer-wise and single-ancilla qubit padding. $Y_{i} := \lceil {{K} \over {2^{i-1}}} \rceil \mod 2$ is an odd number of qubits in the $i\mathrm{th}$ layer when divided by a power of two.}
  \label{tab:table2}
  \begin{tabular}{ccc}
    \toprule
    Padding Methods & Layer-wise qubit padding & Single-ancilla qubit padding \\
    \midrule
    Ancillary qubits & $\sum_{i=1}^{m-1} Y_{i}$ & 1 \\ 
    \midrule
    Circuit depth & $\sum_{i=1}^{m} (2 l_{i} + 1) - l_{m}$ & $\sum_{i=1}^{m} (2 l_{i} + 1) - l_{m}$ \\ 
    \midrule
    Parameter-sharing on & $\sum_{i=1}^{m} (l_{i} + 1)$ & $\sum_{i=1}^{m} (l_{i} + 1)$ \\ 
    \midrule
    Parameter-sharing off & $\sum_{i=1}^{m-1} ( (\lceil {{K} \over {2^{i-1}}} \rceil + Y_{i} ) (l_{i} + {1 \over 2} )) + l_{m} + 1 $ & $\sum_{i=1}^{m-1} ( (\lceil {{K} \over {2^{i-1}}} \rceil + Y_{i} ) (l_{i} + {1 \over 2} )) + l_{m} + 1 $ \\ 
  \bottomrule
\end{tabular}
\end{table}

\begin{table}[h]
  \caption{Comparison of the number of ancillary qubits, circuit depth, and total number of parameters for each padding method with different ansatz sets based on five initial input qubits.}
  \label{tab:table3}
  \resizebox{\textwidth}{!}{
  \begin{tabular}{cccccccc}
    \toprule
    Padding Methods & Ancillary qubits & Circuit depth & \multicolumn{2}{c}{Parameter-sharing on} & \multicolumn{2}{c}{Parameter-sharing off} \\
                    & & & ansatz set 1 & ansatz set 2 & ansatz set 1 & ansatz set 2 & \\
    \midrule
    Classical data padding & 3 & 8 & 12 & 45 & 40 & 195 \\ 
    Skip pooling & 0 & 10 & 12 & 45 & 26 & 135 \\ 
    Layer-wise qubit padding & 2 & 8 & 12 & 45 & 34 & 165 \\ 
    Single-ancilla qubit padding & 1 & 8 & 12 & 45 & 34 & 165\\ 
  \bottomrule
\end{tabular}
}
\end{table}

In this section, we present numerical experimental results that evaluate the performance of QCNNs with various padding methods for binary classification tasks in a noiseless environment. Tables~\ref{tab:table1} and~\ref{tab:table2} summarize the number of ancillary qubits, circuit depth, and number of parameters required for the naive and proposed methods. Classical data padding and skip pooling methods exhibit a significant difference in terms of the utilization of ancillary qubits. Specifically, classical data padding maximally uses ancillary qubits to apply a natural QCNN algorithm, whereas skip pooling does not use any ancillary qubits. However, skip pooling poses a potential drawback in the form of a potential increase in circuit depth, which affects computational power and runtime. Additionally, the use of fewer qubits results in the use of fewer parameters when they are not shared by each layer. However, the qubit padding method can apply an efficient QCNN algorithm with fewer qubits. Because the single-ancilla qubit padding method uses only one ancillary qubit, it does not incur additional circuit depth and offers the advantage of utilizing a slightly larger number of parameters than the skip-pooling method. 

The simulation results were based on experiments using two different ansatz sets, denoted as ansatz set 1 and ansatz set 2. Table~\ref{tab:table3} lists the number of ancillary qubits, circuit depth, and parameters when applying different ansatz sets to the QCNN. All convolutional layers were used only once, and without loss of generality, the circuit depth was obtained by setting the convolution and pooling gate depths to 1. We obtained results from 10 repeated experiments with randomly initialized parameters for each ansatz set. The performance of the QCNN model was evaluated using the mean squared error (MSE) loss function. Model parameters were updated using the Adam optimizer ~\cite{kingma2014adam}. The learning rate and batch size were set to 0.01 and 25, respectively. Training was performed with 10 and 50 epochs on the MNIST and Breast Cancer datasets, respectively.

\begin{table}[h]
\centering
  \caption{QCNN model performance with various padding methods constructed using ansatz set 1. Mean accuracy and standard deviation of classification for 0 \& 1 and 5 \& 6 in the MNIST dataset, as well as whether tumors were malignant or benign in the Breast Cancer dataset. The mean and standard deviation were obtained from 10 repeated experiments with parameters initialized randomly.
  }
  \label{tab:table4}
  \begin{tabular}{cccccc}
    \toprule
    Padding Methods         & Parameters    & MNIST 0 \& 1          & MNIST 5 \& 6          & Breast Cancer         \\
    \midrule
    Zero-data padding       & Shared        & 0.9127 $\pm$ 0.0283   & 0.8040 $\pm$ 0.0305   & 0.7085 $\pm$ 0.0548   \\ 
                            & Not shared    & 0.9760 $\pm$ 0.0101   & 0.9175 $\pm$ 0.0168   & 0.8021 $\pm$ 0.0366   \\ 
    \midrule
    Periodic-data padding   & Shared        & 0.9024 $\pm$ 0.0358   & 0.8078 $\pm$ 0.0280   & 0.6851 $\pm$ 0.0764   \\ 
                            & Not shared    & 0.9729 $\pm$ 0.0094   & 0.9247 $\pm$ 0.0092   & 0.7521 $\pm$ 0.0578   \\ 
    \midrule
    Layer-wise ancilla      & Shared        & 0.8982 $\pm$ 0.0410   & 0.8147 $\pm$ 0.0299   & 0.7095 $\pm$ 0.0486   \\ 
                            & Not shared    & 0.9729 $\pm$ 0.0157   & 0.9222 $\pm$ 0.0123   & 0.7723 $\pm$ 0.0236   \\ 
    \midrule
    Single-ancilla          & Shared        & 0.9175 $\pm$ 0.0257   & 0.8254 $\pm$ 0.0454   & 0.7468 $\pm$ 0.0378   \\ 
                            & Not shared    & 0.9859 $\pm$ 0.0059   & 0.9297 $\pm$ 0.0164   & 0.7808 $\pm$ 0.0310   \\ 
    \midrule
    Skip Pooling            & Shared        & 0.9176 $\pm$ 0.0298   & 0.8613 $\pm$ 0.0274   & 0.7659 $\pm$ 0.0563   \\ 
                            & Not shared    & 0.9834 $\pm$ 0.0057   & 0.9347 $\pm$ 0.0121   & 0.7882 $\pm$ 0.0466  \\ 
  \bottomrule
\end{tabular}
\end{table}

\begin{table}[h]
\centering
  \caption{QCNN model performance with various padding methods constructed using ansatz set 2. Mean accuracy and standard deviation of classification for 0 \& 1 and 5 \& 6 in the MNIST dataset, as well as whether tumors were malignant or benign in the Breast Cancer dataset. The mean and standard deviation were obtained from 10 repeated experiments with parameters initialized randomly.
  }
  \label{tab:table5}
  \begin{tabular}{cccccc}
    \toprule
    Padding Methods         & Parameters    & MNIST 0 \& 1          & MNIST 5 \& 6          & Breast Cancer         \\
    \midrule
    Zero-data padding       & Shared        & 0.9806 $\pm$ 0.0189   & 0.9142 $\pm$ 0.0134   & 0.7399 $\pm$ 0.0510   \\ 
                            & Not shared    & 0.9962 $\pm$ 0.0014   & 0.9666 $\pm$ 0.0045   & 0.8095 $\pm$ 0.0373   \\ 
    \midrule
    Periodic-data padding   & Shared        & 0.9749 $\pm$ 0.0097   & 0.9081 $\pm$ 0.0128   & 0.7138 $\pm$ 0.0559   \\ 
                            & Not shared    & 0.9964 $\pm$ 0.0013   & 0.9668 $\pm$ 0.0058   & 0.7850 $\pm$ 0.0483   \\ 
    \midrule
    Layer-wise ancilla      & Shared        & 0.9782 $\pm$ 0.0060   & 0.9248 $\pm$ 0.0153   & 0.7425 $\pm$ 0.0324   \\ 
                            & Not shared    & 0.9952 $\pm$ 0.0029   & 0.9566 $\pm$ 0.0087   & 0.8021 $\pm$ 0.0429   \\ 
    \midrule
    Single-ancilla          & Shared        & 0.9886 $\pm$ 0.0048   & 0.9307 $\pm$ 0.0114   & 0.7712 $\pm$ 0.0617   \\ 
                            & Not shared    & 0.9962 $\pm$ 0.0027   & 0.9669 $\pm$ 0.0072   & 0.8122 $\pm$ 0.0358   \\ 
    \midrule
    Skip Pooling            & Shared        & 0.9903 $\pm$ 0.0061   & 0.9459 $\pm$ 0.0062   & 0.7744 $\pm$ 0.0368   \\ 
                            & Not shared    & 0.9969 $\pm$ 0.0016   & 0.966 $\pm$ 0.0034    & 0.8155 $\pm$ 0.0311  \\ 
  \bottomrule
\end{tabular}
\end{table}

Table~\ref{tab:table4} lists the average accuracies and standard deviations of the binary classification task on the MNIST (labels 0 \& 1 and 5 \& 6) and Breast Cancer datasets obtained using ansatz set 1. Both single-ancilla qubit padding and skip pooling achieved superior accuracy across the two datasets. For example, for labels 0 \& 1 in the MNIST dataset, single-ancilla qubit padding with shared parameters achieved an average accuracy of 0.9175 ($\pm$ 0.0257), whereas skip pooling showed an accuracy of 0.9176 ($\pm$ 0.0298). Such a trend was consistently observed in all other test cases, indicating the effectiveness of these methods in handling classification tasks with greater precision. However, the other padding methods, although still effective, did not attain the same levels of accuracy as single-ancilla qubit padding and skip pooling. Table~\ref{tab:table5} lists results obtained using ansatz set 2. Single-ancilla qubit padding and skip pooling consistently achieved higher performance. For example, in the Breast Cancer dataset, single-ancilla qubit padding achieved an average accuracy of 0.7712 ($\pm$ 0.0617) and skip pooling achieved 0.7744 ($\pm$ 0.0368), showing better results with fewer qubits. Although skip pooling is efficient in terms of qubit usage and performance, it results in a deeper circuit that may be more susceptible to noise, particularly in existing noisy quantum devices. In contrast, single-ancilla qubit padding is more robust to noise, potentially making it more suitable for implementation on quantum devices. This will be demonstrated in the following section.

\subsection{Simulation with noise}

We conducted an additional experiment to evaluate the impact of noise in a quantum computing environment on the performance of QCNN algorithms. In particular, we considered the influence of circuit depth on error accumulation in quantum computation by comparing performance between the single-ancilla qubit padding and skip pooling methods. The noise simulations focused on types of noise that closely relate to circuit depth, and state preparation and measurement errors (SPAM) were excluded, as they were beyond the scope of our interest in this study. We considered various types of noise in quantum devices, such as depolarization errors, gate lengths, and thermal relaxation, but not the physical connectivity of qubits. Appendix ~\ref{app:appendix_b} provides details of the noise circuits used in this experiment.

\begin{table}
  \caption{Average error rates for the IBMQ device, \textit{JaKarta} used in this experiment.
  }
  \label{tab:table6}
  \resizebox{\textwidth}{!}{
  \begin{tabular}{cccccc}
    \toprule
    1-qubit Depolarizing & 2-qubit Depolarizing & 1-qubit Gate Length & 2-qubit Gate Length & $T_{1}$ & $T_{2}$ \\
    \midrule
    0.0004 & 0.0126 & 35.56 (ns) & 327.11 (ns) & 128.43 (us)& 33.85 (us)   \\ 
  \bottomrule
\end{tabular}
}
\end{table}

\begin{figure}[ht]
  \centering
    \begin{subfigure}[b]{0.33\textwidth}
        \includegraphics[width=1\linewidth]{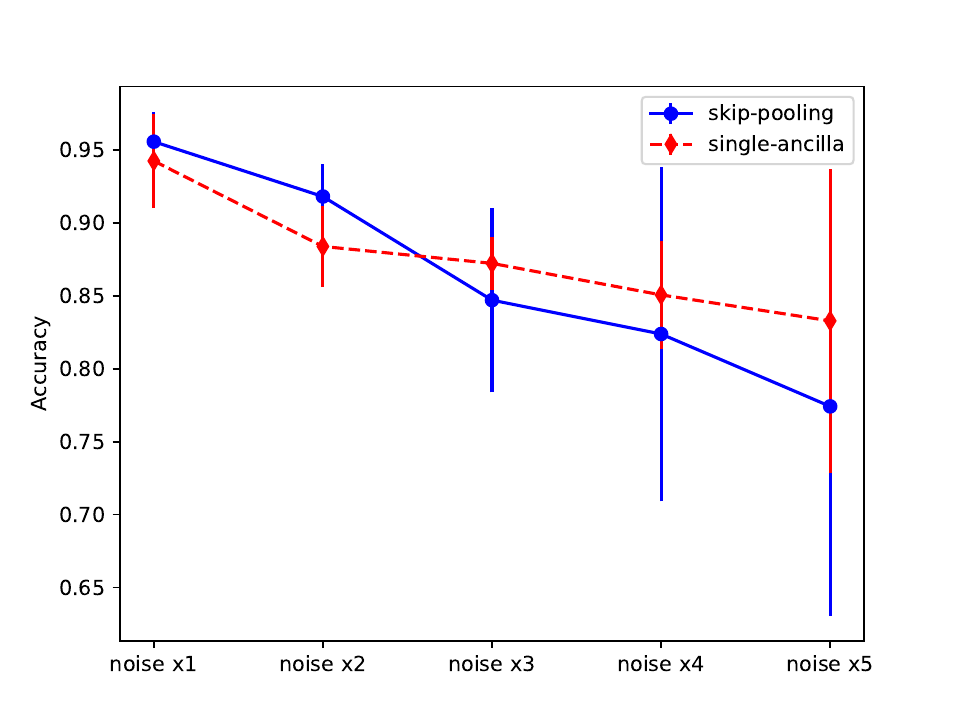}
        \caption{MNIST 0 \& 1}
        \label{fig:fig6_1}
    \end{subfigure}%
    \begin{subfigure}[b]{0.33\textwidth}
        \includegraphics[width=1\linewidth]{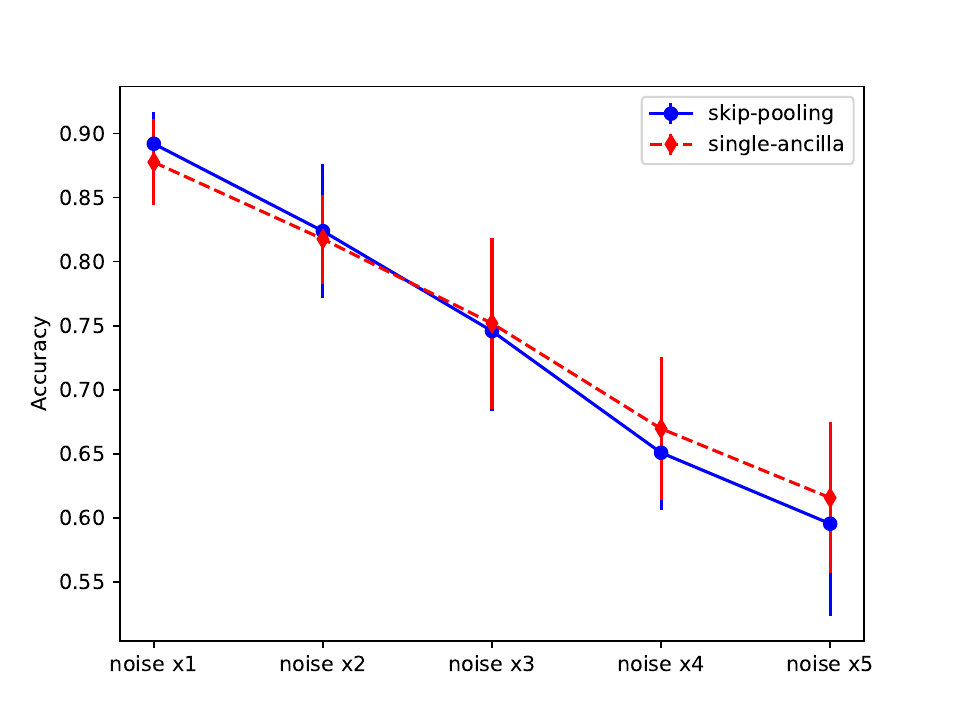}
        \caption{MNIST 5 \& 6}
        \label{fig:fig6_2}
    \end{subfigure}%
    \begin{subfigure}[b]{0.33\textwidth}
        \includegraphics[width=1\linewidth]{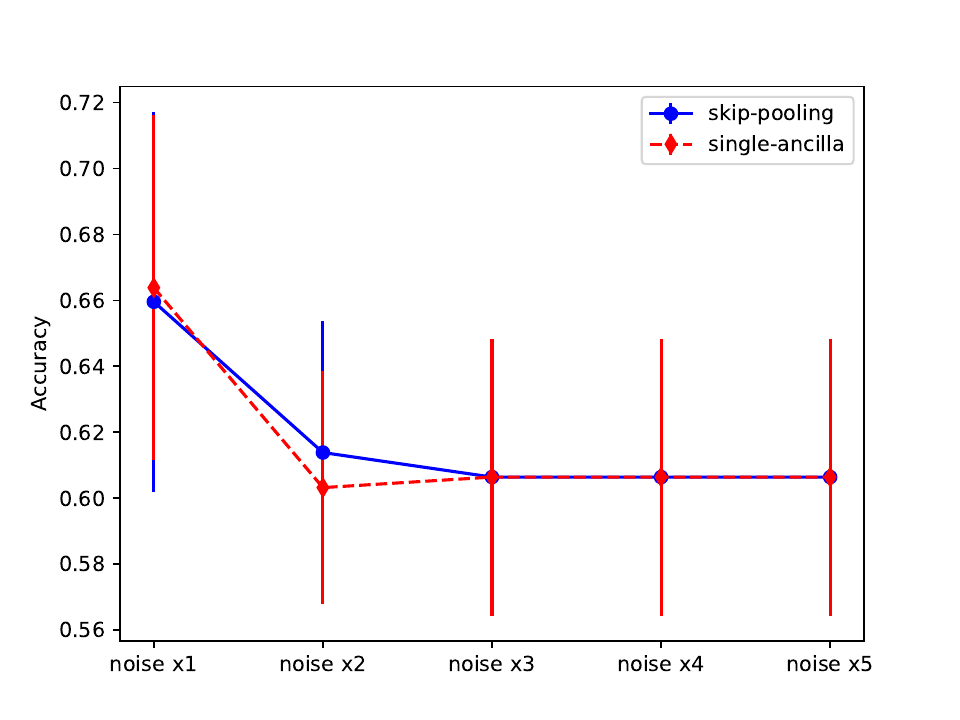}
        \caption{Breast cancer}
        \label{fig:fig6_3}
    \end{subfigure}%
    \caption{Test results of QCNN model with skip pooling and single-ancilla qubit padding constructed with ansatz set 2. The depolarizing errors and gate lengths were increased from their original values (x1) up to a maximum of (x5). The solid blue line represents skip pooling, whereas the dashed red line denotes single-ancilla qubit padding. The mean and standard deviation were obtained from 10 repeated experiments with parameters initialized randomly. (a) shows the average accuracy and standard deviation of classification for 0 \& 1 in the MNIST dataset, (b) shows the average accuracy and standard deviation of classification for 5 \& 6 in the MNIST dataset, and (c) shows the average accuracy and standard deviation of classification for the Breast Cancer dataset. 
  }
  \label{fig:fig6}
\end{figure}

We used the noise parameters observed from IBMQ Jakarta, a real quantum device, to simulate a realistic noise model. Table~\ref{tab:table6} lists the average error rates observed on IBMQ Jakarta. To evaluate the influence of a range of noise levels, we performed experiments with depolarizing errors and gate lengths ranging from one to five times the original values. This multiplication was applied consistently over 10 repeated experiments with randomly initialized parameters, and the results are shown in Figure~\ref{fig:fig6}. For the MNIST dataset, as the noise level increased, single-ancilla qubit padding exhibited a smaller degradation in accuracy than skip pooling. For the Breast Cancer dataset, performance did not change significantly as noise increased, as the label ratio was approximately 6:4. These results demonstrate that the proposed method provides an optimal solution for constructing an efficient QCNN architecture with minimal resource overhead, maintaining robust performance despite noise and imperfections.

\section{Extension to Multi-Qubit Quantum Convolutional Operations}
\label{sec:section5}
An arbitrary unitary operation acting on $n$ qubits, which is an element in the $SU(2^n)$ group, can be specified using $4^n-1$ real parameters. This implies that the number of elementary gates required to implement an arbitrary $n$-qubit unitary operation increases exponentially with $n$. Therefore, minimizing the number of qubits involved in a quantum convolutional operation is beneficial in practice. This is a primary motivation for designing quantum convolutional operations that act on only two qubits, as considered in this study. However, quantum convolutional operations can theoretically act on any $n$ number of qubits. In general, the quantum circuit depth of any given convolutional layer, denoted by $l$, is greater than or equal to $n$ (i.e., $l\ge n$), and equality is satisfied only when the quantum convolutional layer consists of $an$ qubits where $a\in\mathbb{Z}_+$ is a positive integer (i.e., $m$ is a positive-integer multiple of $n$). Therefore, if the number of qubits in a quantum convolutional layer, denoted by $m$, is not an integer multiple of $n$, the circuit depth of the given convolutional layer can be minimized at the cost of introducing $n' < n$ ancilla qubits such that $m+n'= an$. 

For example, consider a quantum convolutional layer consisting of $m=7$ qubits where each quantum convolutional operation acts on $n=3$ qubits. By utilizing $n'=2$ ancilla qubits, the circuit depth of the quantum convolutional layer can be minimized to three.

\section{Conclusions}
\label{sec:section6}
In this study, we designed a QCNN architecture that can handle arbitrary data dimensions. Using qubit padding, we optimized the allocation of quantum resources through the efficient use of ancillary qubits. Our method not only reduces the number of ancillary qubits, but also optimizes the circuit depth to construct an efficient QCNN architecture. This results in an optimal solution that is computationally efficient and robust against noise. We benchmarked the performance of our QCNN using both naive methods and the proposed methods on the MNIST and Breast Cancer datasets for binary classification. In simulations without noise, both skip pooling and our proposed single-ancilla qubit padding method achieved high accuracy in most cases. We also compared performance between single-ancilla qubit padding and skip pooling in a noisy simulation, using the noise model and parameters of an IBM quantum device. Our results demonstrate that as the noise level increases, single-ancilla qubit padding exhibits less performance degradation and lower sensitivity to variation. Therefore, the proposed method serves as a fundamental building block for the effective application of QCNN to real-world data of arbitrary input dimension.

\section*{Acknowledgments}
This research was supported by Institute for Information \& Communications Technology Promotion (IITP) grant funded by the Korea government (No. 2019-0-00003, Research and Development of Core Technologies for Programming, Running, Implementing and Validating of Fault-Tolerant Quantum Computing System), the Yonsei University Research Fund of 2023 (2023-22-0072), the National Research Foundation of Korea (Grant No. 2022M3E4A1074591), and the KIST Institutional Program (2E32941-24-008).

\appendix
\setcounter{section}{0}
\renewcommand{\thesection}{\text{Appendix }\Alph{section}} 
\renewcommand\thefigure{\Alph{section}.\arabic{figure}} 
\renewcommand\theequation{\Alph{section}.\arabic{equation}} 
\renewcommand\thetable{\Alph{section}.\arabic{table}} 

\section{Quantum Gates Definition}
\label{app:appendix_a}
The single qubit rotation about the Y and Z axis of the Bloch sphere is defined as follows: 
\begin{align*}
    R_{Y}(\theta) = \exp{\left( -i {\theta \over 2 } Y \right)}  = \begin{pmatrix} \cos{\frac{\theta}{2}} & -\sin{\frac{\theta}{2}} \\ \sin{\frac{\theta}{2}} & \cos{\frac{\theta}{2}} \end{pmatrix}, \quad
    R_{Z}(\lambda) = \exp{\left( -i {\lambda \over 2 } Z \right)}  = \begin{pmatrix} e^{ -i {\lambda \over 2 } } & 0 \\ 0 & e^{ i {\lambda \over 2 }} \end{pmatrix}. 
\end{align*}
The SX gate, which is the square root of the single-qubit X gate, is defined as follows:
\begin{align*}
    SX = \sqrt{X} = {1 \over 2} \begin{pmatrix} 1 + i & 1 - i \\ 1 - i & 1 + i \end{pmatrix}.
\end{align*}
Morevoer, the arbitrary single qubit unitary $U_{3}$ is defined as follows:
\begin{align*}
    U_3(\theta, \phi, \lambda) = \begin{pmatrix} \cos{\frac{\theta}{2}} & -e^{i\lambda}\sin{\frac{\theta}{2}} \\e^{i\phi}\sin{\frac{\theta}{2}} & e^{i(\phi+\lambda)}\cos{\frac{\theta}{2}}
    \end{pmatrix}.
\end{align*}
The two-qubit controlled operation $CU$ is defined as follows:
\begin{align*}
    CU = |0 \rangle\!\langle 0| \otimes I + |1 \rangle\!\langle 1| \otimes U.
\end{align*}
For the $CX$ gate, $U$ becomes the $X$ gate, which is a single-qubit bit-flip gate. For the $CRY$ and $CRZ$ gates, $U$ represents the single-qubit rotation around the $y$ or $z$ axis of the Bloch sphere corresponding to the $RY$ or $RZ$ gate, respectively.

\section{Noise model}
\label{app:appendix_b}
\subsection{Quantum Noise Channels}
In this section, we introduce three types of errors that are prevalent in quantum computing: (1) depolarizing errors, (2) thermal relaxation errors, and (3) state preparation and measurement errors (SPAM).

\subsubsection{Depolarizing errors}
Depolarizing channels are simple noise models used in quantum systems. If a single qubit is depolarized with probability $p$, it is completely replaced by a mixed state ${I \over 2}$. The qubit is left untouched with probability $1 - p$. The resulting quantum state is
\begin{align*}
    \mathcal{E}(\rho) = {pI \over 2} + (1 - p) \rho.
\end{align*}
Depolarizing channels can be extended to quantum systems in more than two dimensions. In the case of a quantum system with $d$ dimensions, the depolarizing channel replaces the quantum state with a completely mixed state ${I \over d}$ with a probability of $p$, and leaves the state untouched otherwise. The quantum operation is
\begin{align*}
    \mathcal{E}(\rho) = {pI \over d} + (1 - p) \rho.
\end{align*}
Let us now consider the case of a single qubit. Assuming that three types of errors occur with probability $p_{1}$ - bit flips, phase flips, and both bit and phase flips - the depolarizing channel can be represented by the following operator:
\begin{align*}
    \rho \to \mathcal{E}_{D}(\rho)  = \sum_{i=0}^{3} K_{i} \rho K_{i}^{\dagger},	\\
\end{align*}
\begin{align*}
    K_{0} = \sqrt{1 - p_{1}} I, K_{1} = \sqrt{{p_{1} \over 3}} X, K_{2} = \sqrt{{p_{1} \over 3}} Y, K_{3} = \sqrt{{p_{1} \over 3}} Z,
\end{align*}
where $\rho$ denotes the density matrix of the qubit.

\subsubsection{Thermal relaxation errors}
Thermal relaxation is a phenomenon wherein a system prepared in a higher-energy state spontaneously releases energy over time and settles into a ground state owing to interactions with its environment. This phenomenon is governed by the relaxation times $T_{1}$ and $T_{2}$, which indicate the time required for the system to decay from the excited state to the ground state and that required for qubit dephasing, respectively. We also have the average execution time for each type of quantum gate $g$, denoted $T_{g}$. A reset error is likely to occur, and the weights that determine which of the two equilibrium states ($|0 \rangle$ or $|1 \rangle$) it tends toward, called the excited-state population $(0 \le p_{e} \le 1)$, can be calculated as follows:
\begin{align*}
    p_{e} = \left( 1 + \mathrm{exp} \left\{ {{2hf} \over {k_{B} T}} \right\}  \right)^{-1},
\end{align*}
where $T$ is the temperature in \si{\kelvin}, $f$ is the qubit frequency in \si{\hertz}, $k_{B}$ is Boltzmann’s constant (\si{\electronvolt\per\kelvin}), and $h$ is Planck’s constant (\si{\electronvolt\second}). If the frequency $f \to \infty$ or temperature $T \to 0$, the excited state population is zero. As $T = 0$ in our noise model, we consider $p_{e} = 0$. We can define $\epsilon_{T_{1}} = \mathrm{exp} \left\{ - {{T_{g}} / {T_{1}}} \right\}$ and $\epsilon_{T_{2}} = \mathrm{exp} \left\{ - {{T_{g}} / {T_{2}}} \right\}$ as the error rates at which relaxation and dephasing will be applied after the gate is applied to each qubit. The reset probability can then be defined as $p_{reset} = 1 - \epsilon_{T_{1}}$.

If $T_{2} \le T_{1}$, the thermal relaxation channel is a probabilistic mixture of the reset operations and unitary errors, which can be expressed as
\begin{align*}
    p_{z} &= (1 - p_{reset}) (1 - \epsilon_{T_{2}} / \epsilon_{T_{1}}) / 2 ,\\
    p_{r0} &= (1 - p_{e}) p_{reset} ,\\
    p_{r1} &= p_{e} p_{reset} ,\\
    p_{id} &= 1 - p_{z} - p_{r0} - p_{r1}.
\end{align*}
This can be represented using a Kraus matrix:
\begin{align*}
    K_{0} = \sqrt{1 - p_{z} - p_{r0} - p_{r1}} \begin{bmatrix} 1 & 0 \\ 0 & 1 \end{bmatrix}
    , K_{1} = \sqrt{p_{z}} \begin{bmatrix} 1 & 0 \\ 0 & -1 \end{bmatrix} 
    , K_{2} = \sqrt{p_{r0}} \begin{bmatrix} 1 & 0 \\ 0 & 0 \end{bmatrix},\\
    K_{3} = \sqrt{p_{r0}} \begin{bmatrix} 0 & 1 \\ 0 & 0 \end{bmatrix}
    , K_{4} = \sqrt{p_{r1}} \begin{bmatrix} 0 & 0 \\ 1 & 0 \end{bmatrix}
    , K_{5} = \sqrt{p_{r1}} \begin{bmatrix} 0 & 0 \\ 0 & 1 \end{bmatrix},
\end{align*}
where $p_{r0} \in [0, 1]$ is the probability of a reset to 0, $p_{r1} \in [0, 1]$ is the probability of a reset to 1, and $p_{z} \in [0, 1]$ is the probability of a phase-flip (Pauli Z) error. 

If $T_{2} > T_{1}$, the thermal relaxation channel can be described by the Choi-matrix representation. The Choi matrix is expressed as
\begin{align*}
    \Lambda =   \begin{bmatrix}
                1 - p_{e} p_{reset} & 0 & 0 & \epsilon_{T_{2}} \\
                0 & p_{e} p_{reset} & 0 & 0 \\
                0 & 0 & (1 - p_{e}) p_{reset} & 0 \\
                \epsilon_{T_{2}} & 0 & 0 & 1 - (1 - p_{e}) p_{reset} 
                \end{bmatrix}.
\end{align*}

\subsubsection{State preparation and measurement errors}
SPAM errors can occur during the preparation (initialization) of a quantum state, and during the measurement of the final quantum state. In IBM Quantum, the initial state is $|0\rangle$. However, we can start from a different initial state, with the computational process used to run the algorithm to obtain this initial state leading to errors. The probability of measurement error is defined as $P(i|j)$, where $j$ is the true ideal measurement outcome, $i$ is the noisy measurement outcome, and $i$ and $j$ are the integer representations of bit strings. Because we were only interested in circuit depth, we did not consider SPAM errors as part of the main run in our noise model.

\subsubsection{Noise simulation}

\begin{figure}[ht]
  \centering
  \includegraphics[width=\linewidth]{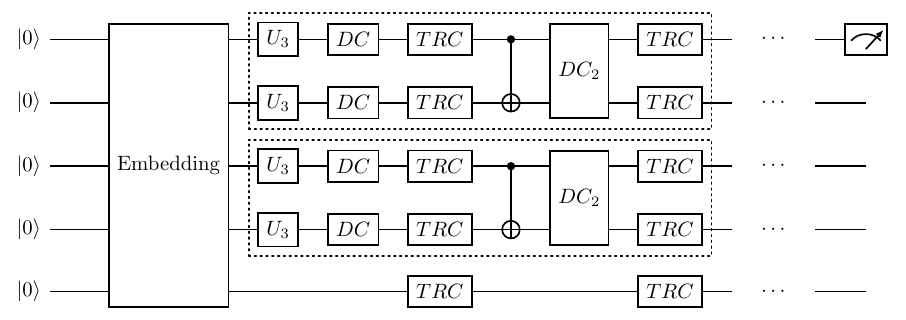} 
  \caption{Example of noise model with a five-qubit QCNN structure. The depolarizing channel (DC) is applied after every gate. The thermal relaxation channel (TRC) is applied after the DC. Because other qubits are also exposed to thermal relaxation time during gate operations, the TRC is applied equally to all qubits. We ignored SPAM errors, which occur during data embedding and measurement.
  }
  \label{fig:fig7}
\end{figure}

In quantum devices, directly executed circuits contain single- and two-qubit gates. The gate in our experiment was compiled into the basis gate set of quantum hardware, including the identity gate (ID), rotation-Z gate (RZ), root-X gate (SX), X gate (X), U3 gate (U3), and two-qubit controlled-X gate (CX). Single- and two-qubit gate errors consist of a depolarizing channel followed by a thermal relaxation channel. All quantum channels were applied independently, and their combination was computed by constructing error operators on the circuit gates. The experimental circuit was executed on PennyLane using the noise information obtained from an IBM quantum device~\cite{bergholm2018pennylane}. Figure~\ref{fig:fig7} illustrates a depolarizing channel and a thermal relaxation channel applied to single- and two-qubit gates in a five-qubit quantum circuit.

\end{document}